\documentclass[12pt]{article}
\usepackage{graphicx}
\usepackage[body={6.9in, 9.0in},left=1.0in, top = 1.35in]{geometry}
\usepackage{epsfig}
\usepackage{appendix}
\begin{document}
\title{Isgur-Wise function in a QCD inspired potential model with  WKB Approximation }
\author{$^{1}$Bhaskar Jyoti Hazarika and $^{2}$D K Choudhury \\
$^{1}$Dept of Physics,Pandu College,Guwahati-781012,India\\
e-mail:bh53033@gmail.com\\
$^{2}$Dept. of Physics, Gauhati University, Guwahati-781014,India}
\date{}
\maketitle
\begin{abstract}
We use WKB Approximation in the calculation of slope and curvature of Isgur-Wise function in a QCD inspired potential model.This work is an extension of the approximation methods to the QCD inspired potential model.The approach hints at an effective range of distance for the calculation of slope and curvature of Isgur-Wise function.Comparision is also made with those of Dalgarno and VIPT to see the advantages of using WKB Approximation.\\

\end{abstract}
\section{Introduction}
Probably this is the first attempt to use WKB Approximation (WKBA) for the calculation of slope and curvature of Isgur-Wise (IW) function with linear cum Coulombic pottential of QCD.The WKBA which gives approximate but direct solution of Schr\"{o}dinger equation is applicable to regions where potential energy variation is small.There are few works \cite{1,2,3}, where linear cum Coulomb potential is used under WKBA to calculate the energy spectrum ,decay width etc.With the success of those models ,we here try to extend WKBA in the calculation of slope and curvature of IW function  to sort out the advantages as well as the limitations of the method.\\
It is observed that so far the results for slope and curvature are not satisfactory for $B_{c}$ meson [4-10]in our QCD inspired potential model with significant confinement effect $b=0.183GeV^{2}$ \cite{11}.We opt this method in the hope of doing better for the  $B_{c}$ meson.\\

\section{Formalism}
The one dimensional(say $x$)  Schr\"{o}dinger equation is \cite{19}:
\begin{equation}
\frac{d^2\Psi}{dx^2}+2\mu\left[E-V(x)\right]\Psi=0
\end{equation}
For $E>V(x)$ ,it becomes :
\begin{equation}
\frac{d^2\Psi}{dx^2}+k^{2}(x)\Psi=0
\end{equation}
where
\begin{equation}
k^{2}(x)=2\mu\left[E-V(x)\right]
\end{equation}
The WKB solution of Eq.(2) is :
 \begin{equation}
\Psi(x)=\frac{N}{\sqrt{ k(x)}}Cos(\int  k(x) dx-\frac{\pi}{4})
\end{equation}
 For $E<V(x)$ , the WKB solution is either:
\begin{equation}
\Psi(x)=\frac{N^{\prime}}{\sqrt{k^{\prime}(x)}}e^{-\int  k^{\prime}(x) dx} 
\end{equation}
or
\begin{equation}
\Psi(x)=\frac{N^{\prime\prime}}{\sqrt{k^{\prime}(x)}}e^{\int  k^{\prime}(x) dx} 
\end{equation}
where
\begin{equation}
k^{\prime^{2}}(x)=2\mu\left[V(x)-E\right]
\end{equation}
and $N$,$N^{\prime}$ , $N^{\prime\prime}$ are the normalization constants .\\
The WKB quantization condition for the allowed bound state energy is \cite{19}:
\begin{equation}
\left(n_{r}+\frac{1}{2}\right)=\int_{A}^{B}k(x) dx
\end{equation}
where $A$ and $B$ represent  the turning points at which $E=V(x)$ and  $n_{r}\ge0$.\\

For the three dimensional radial  Schr\"{o}dinger equation which contain an additional centrifugal potential $\frac{l(l+1)}{2\mu r^{2}}$ to the original potential $V(r)$,the independent variable $r$ is converted to $x$ to give the three dimensional equation a one dimensional form and in the process we instead of the integral $ \int_{A}^{B} k(x) dx $ are left with the integral given as \cite{19} :

\begin{equation}
\int_{A}^{B}k(r) dr=\int_{A}^{B} \left (E-V(r)-\frac{(l+\frac{1}{2})^{2}}{2\mu r^{2}}\right)^{\frac{1}{2}} dr
\end{equation}
We note that the term  $\frac{l(l+1)}{2\mu r^{2}}$ has changed to  $\frac{(l+\frac{1}{2})^{2}}{2\mu r^{2}}$ and the WKB solution is applicable to radial  Schr\"{o}dinger equation like the one dimensional  Schr\"{o}dinger equation in the variable $x$.\\ 

With the original linear cum Coulombic potential \cite{12}the WKB quantization condition in this case (Eq.9 above) for the ground state ($n_{r}=1,l=0$) is :
\begin{equation}
\left(n+\frac{1}{2}\right)=\int_{A}^{B}\left (E-br+\frac{\alpha}{r}-\frac{1}{4r^{2}}\right)^{\frac{1}{2}} dr
\end{equation}
The limits $A$ and $B$ are the turning points  given by  the positive roots of the cubic equation:
\begin{equation}
br^{3}-Er^{2}-\alpha r+\frac{1}{4}=0
\end{equation}

We also  note that  
\begin{equation}
\alpha=\frac{4\alpha_{s}}{3}
\end{equation}

At the turning points, the WKB solutions become invalid and we find wave functions only in the regions $r<A(E<V),A<r<B(E>V)$ and $r>B(E<V)$.We use the leading order expressions for energy given by Eq.(11) of Ref[2] with $W$,$\mu$ replaced by $E$ and $b$ respectively:
\begin{equation}
E=\left(\frac{3\pi b}{4}\right)^{\frac{2}{3}}\left(n+\frac{3}{2}\right)^{\frac{2}{3}}
\end{equation}
The solutions for the respective regions are \cite{19} :
\begin{equation}
\Psi_{1}(r)=\frac{N_{1}}{2\sqrt{\pi k^{\prime}}r}e^{-\int  k^{\prime} dr} , r<A
\end{equation} 
\begin{equation}
\Psi_{2}(r)=\frac{N_{2}}{2\sqrt{\pi k}r}Cos(\int  k dr-\frac{\pi}{4}) ,A<r<A
\end{equation} 

 \begin{equation}
\Psi_{3}(r)=\frac{N_{3}}{2\sqrt{\pi k^{\prime}}r}e^{\int  k^{\prime} dr} , r>B
\end{equation}

where $N_{1},N_{2}$ and $N_{3}$ are the normalization constants in the respective regions while $ k^{\prime}$ and $k$ are  given respectively by:
 \begin{equation}
 k^{\prime}=\sqrt{2 \mu(V-E)}=\sqrt{2 \mu\left(br^{2}-Er-\frac{\alpha}{r}+\frac{1}{4r^{2}}\right)}
\end{equation}
 and
\begin{equation}
 k =\sqrt{2 \mu(E-V)}=\sqrt{2 \mu\left(Er-br^{2}+\frac{\alpha}{r}-\frac{1}{4r^{2}}\right)}
\end{equation}
As we have seen that there are three wave functions corresponding to the three different regions , so we expect three IW functions also for these regions.\\

The IW function and its slope and curvature can be calculated as \cite{4,20}:
\begin{eqnarray}
\xi\left(y\right)&=& \int_{0}^{+\infty} 4\pi r^{2}\left|\psi\left(r\right)\right|^{2}\cos pr dr\\&=&1-\rho^{2}\left(y-1\right)+ C\left(y-1\right)^{2}+...
\end{eqnarray}
where
\begin{equation}
p^{2}=2\mu^{2}\left(y-1\right)
\end{equation}
We however in this work restrict ourselves to a finite maximum of $r$ (i.e.$r_{max}$) instead of taking it as $\infty$ in Eq.(19).This is because for infinite maximum of $r$ we have diverging results.Consideration of finite maximum of $r$ further hints at the usuable range of $r$ in the calculation of IW function.However,only for the region $r>B$, this consideration comes into existence .The regions can be located from fig.1.\\

The integration involved in different equations (14),(15),(16) lead to the well known elliptic integrals \cite{2}.Instead of using the elliptic functions we have integrated them numerically using mathematica software.\\

We use both the $\overline{MS}$ \cite{4} and $V$-schemes\cite{16,17,18} to see the role of $\alpha_{s}$  in WKBA in calculating the IW function.\\

\section{Calculation and Result}

The turning points $A$,$B$ as well as the $r_{max}$ are listed in table 1.Table 2 shows the slope $\rho^{2}$ and curvature $C$ under $\overline{MS}$ with and without relativistic effect while table 3 shows the same for $V$-scheme.In table 4 we have recorded the predictions of the other models to make comparision with our work.\\

\begin{figure}
\includegraphics[width=3.7in,angle=270]{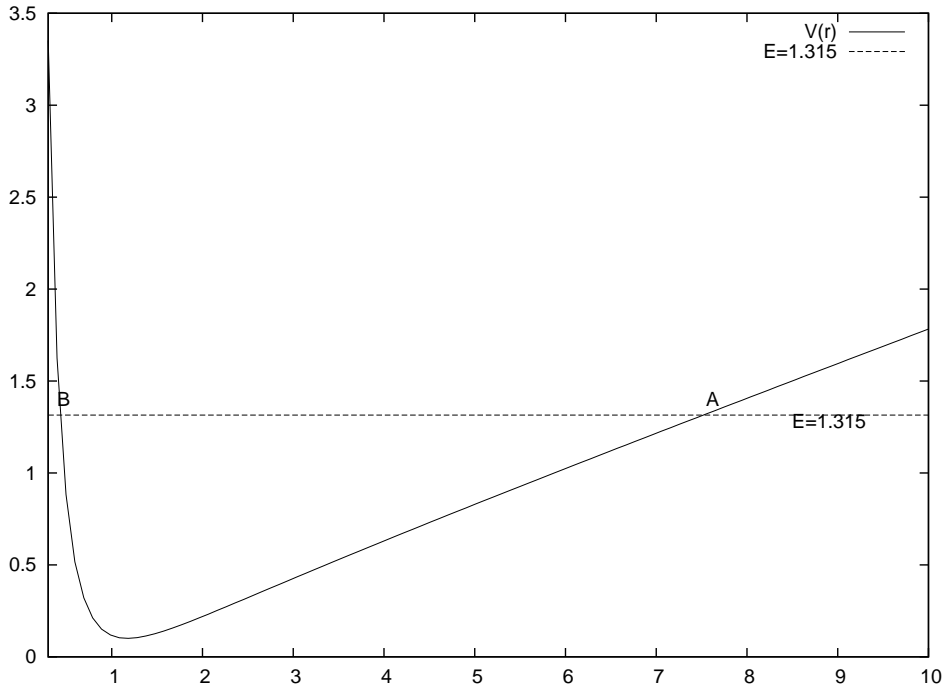}
\caption{Variation of $V(r)$ vs $r$  for the energy $E$ given by Eq.13.The points of intersection between $V(r)$(solid curve) and $E$ (dashed curve) are the turning points $A$ and $B$ -the positive roots of Eq.11. }
\end{figure}

\begin{table}
\begin{center}
\caption{List of the turning points $A$,$B$ in  $GeV^{-1}$.  }

\begin{tabular}{c c c| c c }\hline
Mesons&\multicolumn{2}{c}{$\overline{MS}$ scheme}&\multicolumn{2}{c}{$V$ scheme}\\\cline{2-5}
 &\multicolumn{1}{l}{$A$}&$B$&\multicolumn{1}{l}{$A$}&$B$ \\\hline
$D$&0.431&7.52 &0.337&7.79\\
$D_{s}$&0.353&7.53 &0.269&7.8\\
$B$&0.461&7.36 &0.444&7.4\\
$B_{s}$&0.374&7.375 &0.357&7.41 \\
$B_{c}$&0.195&7.39 &0.182&7.43 \\\hline
\end{tabular}
\end{center}
\end{table}

\begin{table}
\caption{Values of $\rho^{2}$ and $C$ under  $\overline{MS}$-scheme.Subscript `rel' refers to the inclusion of relativistic effect. The maximum finite distance taken for all the mesons are $r_{max}=9.5$  $GeV^{-1}$. }
\begin{tabular}{r| r r r r||r r r r||r r r r}\hline
Me-&\multicolumn{4}{c||}{Region $r<A$}&\multicolumn{4}{c||}{Region $A<r<B$}&\multicolumn{4}{c}{Region $r>B$}\\
\cline{2-13}
son  &\multicolumn{1}{r}{$\rho^{2}$}&$C$&$\rho_{rel}^{2}$&$C_{rel}$ &\multicolumn{1}{r}{$\rho^{2}$}&$C$&$\rho_{rel}^{2}$&$C_{rel}$ &\multicolumn{1}{r}{$\rho^{2}$}&$C$&$\rho_{rel}^{2}$&$C_{rel}$\\\hline 

$D$&\multicolumn{1}{r}{$3.2\times$}&$4.92\times$&$1.2\times$&$1.61\times$&\multicolumn{1}{r}{0.357}&$4.34\times$&0.142&$1.8\times$&\multicolumn{1}{r}{5.09}&0.065&5.06&0.0643\\
&$10^{-3}$&$10^{-6}$&$10^{-3}$ &$10^{-6}$ & &$10^{-3}$ & &$10^{-3}$ & & & &\\
$D_{s}$&\multicolumn{1}{r}{$3.78\times$}&$6.66\times$&$1.51\times$&$2.33\times$&\multicolumn{1}{r}{0.533}&0.012&0.190&$4.3\times$&\multicolumn{1}{r}{9.07}&0.205&9.02&0.2038\\
&$10^{-3}$&$10^{-6}$&$10^{-3}$ &$10^{-6}$& & & &$10^{-3}$ & & &&\\
$B$&\multicolumn{1}{r}{$4.56\times$}&$1.03\times$&$2.8\times$&$5.96\times$&\multicolumn{1}{r}{0.416}&$6.87\times$&0.258&$4.26\times$&\multicolumn{1}{r}{6.41}&0.106&6.38&0.105\\
&$10^{-3}$ &$10^{-5}$  &$10^{-3}$ &$10^{-6}$ & &$10^{-3}$ & &$10^{-3}$ & &&&\\
$B_{s}$&\multicolumn{1}{r}{$5.8\times$}&$1.62\times$&$3.72\times$&$9.74\times$&\multicolumn{1}{r}{0.733}&0.0237&0.436&0.014&\multicolumn{1}{r}{12.6}&0.405&12.5&0.404\\
&$10^{-3}$ &$10^{-5}$ &$10^{-3}$ &$10^{-3}$ & & & & & &&&\\
$B_{c}$&\multicolumn{1}{r}{$1.24\times$}&$6.46\times$&$9.1\times$&$4.4\times$&\multicolumn{1}{r}{3.07}&0.712&1.52&0.353&\multicolumn{1}{r}{90.8}&21.1&90.5&21.0\\
&$10^{-2}$ &$10^{-4}$ &$10^{-3}$ &$10^{-5}$ & & & & & &&&\\\hline
\end{tabular}
\end{table}
\begin{table}
\caption{Values of $\rho^{2}$ and $C$ under  $V$-scheme.Subscript `rel' refers to the inclusion of relativistic effect. The maximum finite distance taken for all the mesons are $r_{max}=9.5$ $GeV^{-1}$.  }
\begin{tabular}{r|r r r r||r r r r||r r r r}\hline
Me-&\multicolumn{4}{c||}{Region $r<A$}&\multicolumn{4}{c||}{Region $A<r<B$}&\multicolumn{4}{c}{Region $r>B$}\\
\cline{2-13}
son  &\multicolumn{1}{r}{$\rho^{2}$}&$C$&$\rho_{rel}^{2}$&$C_{rel}$ &\multicolumn{1}{r}{$\rho^{2}$}&$C$&$\rho_{rel}^{2}$&$C_{rel}$ &\multicolumn{1}{r}{$\rho^{2}$}&$C$&$\rho_{rel}^{2}$&$C_{rel}$\\\hline 

$D$&\multicolumn{1}{r}{$1.87\times$}&$1.62\times$&$1.66\times$&$8.54\times$&\multicolumn{1}{r}{0.296}&$3.76\times$&0.032&$4.07\times$&\multicolumn{1}{r}{5.33}&0.07&5.27&0.07\\
&$10^{-3}$&$10^{-6}$&$10^{-4}$ &$10^{-8}$ & &$10^{-3}$ & &$10^{-4}$ & & & &\\
$D_{s}$&\multicolumn{1}{r}{$2.1\times$}&$1.99\times$&$2.48\times$&$1.4\times$&\multicolumn{1}{r}{0.432}&$9.75\times$&0.036&$8.15\times$&\multicolumn{1}{r}{9.5}&0.215&9.41&0.213\\
&$10^{-3}$ &$10^{-6}$ &$10^{-3}$ &$10^{-7}$& &$10^{-3}$ & &$10^{-4}$ & & &&\\
$B$&\multicolumn{1}{r}{$4.08\times$}&$8.28\times$&$2.26\times$&$4.25\times$&\multicolumn{1}{r}{0.431}&$7.11\times$&0.244&$4.0\times$&\multicolumn{1}{r}{6.46}&0.107&6.43&0.106\\
&$10^{-3}$ &$10^{-6}$ &$10^{-3}$ &$10^{-6}$ & &$10^{-3}$ & &$10^{-3}$ & &&&\\
$B_{s}$&\multicolumn{1}{r}{$5.1\times$}&$1.25\times$&$2.98\times$&$6.73\times$&\multicolumn{1}{r}{0.677}&0.022&0.354&0.011&\multicolumn{1}{r}{12.7}&0.41&12.6&0.407\\
&$10^{-3}$ &$10^{-5}$ &$10^{-3}$ & $10^{-6}$& & & & & &&&\\
$B_{c}$&\multicolumn{1}{r}{$1.2\times$}&$5.7\times$&$8.2\times$&$3.6\times$&\multicolumn{1}{r}{3.18}&0.74&1.32&0.31&\multicolumn{1}{r}{91.5}&21.2&91.1&21.2\\
&$10^{-2}$ &$10^{-5}$ &$10^{-3}$ &$10^{-5}$ & & & & & &&&\\\hline
\end{tabular}
\end{table}

\begin{table}
\begin{center}
\caption{Predictions of the slope and curvature of the IW function in various models.}
\begin{tabular}{c c c}\hline
Model& Value of $\rho^{2}$ &Value of curvature $C$\\\hline
Le Youanc et al \cite{21}&$\ge 0.75$&..\\
Le Youanc et al \cite{22}&$\ge 0.75$&$\ge 0.47$\\ 
Rosner \cite{23}&1.66&2.76\\
Mannel \cite{24,25}&0.98&0.98\\
Pole Ansatz \cite{26}&1.42&2.71\\
MIT Bag Model \cite{28}&2.35&3.95\\
Simple Quark Model \cite{27}&1&1.11\\
Skryme Model \cite{25}&1.3&0.85\\
QCD Sum Rule \cite{26}&0.65&0.47\\
Relativistic Three Quark Model \cite{24}&1.35&1.75\\
Infinite Momentum Frame Quark Model \cite{23}&3.04&6.81\\
Neubert \cite{33}&0.82$\pm$0.09&..\\
UKQCD \cite{32}&$0.83_{-11-22}^{+15+24}$ &..\\
CLEO \cite{33}&$0.76\pm0.16\pm0.08$ &..\\
BELLE \cite{34}&$0.69\pm0.14$&...\\\hline
\end{tabular}
\end{center}
\end{table}

\section{Discussion and Conclusion}

The method used for the first time in the calculation of IW function results in slope and curvature which is comparable with other models for the region $A<r<B$ only.For the other two regions $r<A$ and $r>B$  the results are far from being satisfactory.For the region $r<A$ ($r>B$) the values of $\rho^{2}$ and $C$ are quite small (large) which shows that the IW function is dependent on the distance($r$) considered and the slope and curvature increase with $r$.Further, our assumption of finite maximum ($r_{max}$) distance  instead of infinity as upper limit  ($r=\infty$) is also valid because with increase in $r$ the slope and curvature of IW function increase to great extent leading to quite unsatisfactory result as stated above.Consideration of only the leading term of Eq.11 of Ref[2] may be a cause of these results and hence exploration of approximate expression for energy (Eq.12 of Ref[2]) is another prospect in the model.\\

We have also noticed that   $\rho^{2}$ and $C$ decrease with the increase in $\alpha_{s}$ as our results are larger with  $\overline{MS}$-scheme which have smaller $\alpha_{s}$  than $V$-scheme having larger $\alpha_{s}$.Further, the relativistic effect has reduced the values of  $\rho^{2}$ and $C$ as observed earlier [4-10].However,it is interesting to observe that for the region $r>B$, there is a very little effect of $\alpha_{s}$ as well as  relativistic consideration introduced through $\epsilon$ hinting at the necessity of an effective finite distance as maximum for the calculation of IW function .\\

To conclude , this work opens the idea of a usable range of distance in the calculation of IW function and this can be the alternate consideration that one may take account in renovating the earlier work.\\
      
\end{document}